\documentclass[10pt,letterpaper]{article}
\usepackage{opex3}
\usepackage{amsmath}    
\usepackage{amsfonts}
\usepackage{bm}
\usepackage{amssymb}
\usepackage{appendix}
\usepackage{graphicx}   
\usepackage{subfigure}  
\usepackage{cite}
\usepackage{color}
\usepackage{bigints}

\newcommand{\re}{\textrm{Re}}
\newcommand{\im}{\textrm{Im}}

\newcommand{\dwST}{\delta \omega_{\textrm{ST}}}
\newcommand{\dwPick}{\delta \omega_{\textrm{N-SALT}}}
\newcommand{\dwCS}{\delta \omega_{\textrm{CS}}}
\newcommand{\dwFDTD}{\delta \omega_{\textrm{FDTD}}}
\newcommand{\Half}{\frac{1}{2}}

\begin{document}

\title{Quantitative test of general theories of the intrinsic laser linewidth}

\author{Alexander Cerjan$^1$, Adi Pick$^2$, Y.~D.~Chong$^3$, Steven G.~Johnson$^4$, and A.~Douglas Stone$^{1,*}$} 
\address{$^1$Department of Applied Physics, Yale University, New Haven, Connecticut 06520, USA \newline 
$^2$Department of Physics, Harvard University, Cambridge, Massachusetts 02138, USA\newline 
$^3$School of Physical and Mathematical Sciences, Nanyang Technological University, Singapore 637371, Singapore \newline
$^4$Department of Mathematics, Massachusetts Institute of Technology, Cambridge, Massachusetts 02139, USA}



\date{\today}

\begin{abstract}
We perform a first-principles calculation of the quantum-limited laser
linewidth, testing the predictions of recently developed theories of
the laser linewidth
based on fluctuations about the known steady-state laser solutions against
traditional forms of the Schawlow-Townes linewidth.
The numerical study is based on
finite-difference time-domain simulations of the semiclassical
Maxwell-Bloch lasing equations, augmented with Langevin force terms, and
includes the effects of dispersion, losses due to the open boundary of the
laser cavity, and non-linear coupling between the amplitude and phase fluctuations ($\alpha$ factor).
We find quantitative agreement between the numerical results and the
predictions of the noisy steady-state \textit{ab initio} laser theory (N-SALT), both in the variation
of the linewidth with output power, as well as the emergence of
side-peaks due to relaxation oscillations.
\end{abstract}

\ocis{(140.3430) Laser Theory; (140.3945) Microcavities.} 


\section{Introduction}

The most important property of lasers not captured by semiclassical theories, which treat the
fields via Maxwell's equations, is the intrinsic laser linewidth due to quantum fluctuations.
Above the laser threshold these fluctuations cause a diffusion in the phase of the emitted
laser signal, leading to a broadening of the observed line, which would have zero width within semiclassical
theory. The magnitude of this linewidth depends upon the geometry of the laser cavity as well as upon 
the output power of the laser, and was first calculated by Schawlow and Townes \cite{schawlow_1958}, and the standard formula arising from their work, the ``Schawlow-Townes'' (ST) linewidth, is
\begin{equation}
\dwST = \frac{\hbar \omega_0 \gamma_c^2}{2 P}
\end{equation}
where $\omega_0$ is the central frequency of the emitted laser light, $\gamma_c$ is the
decay rate of the passive cavity resonance corresponding to the laser mode,
and $P$ is the output power. (Schwalow and Townes actually found twice this value in their original work, which assumed the 
laser was near threshold, but it was quickly recognized that far above threshold only the phase fluctuations were important, 
reducing the linewidth by a factor of two). In subsequent decades, improved theoretical analyses allowed for the discovery of 
four significant corrections to this formula.  
The $\alpha$ factor arises from the coupling between intensity and phase fluctuations, and takes different forms depending on 
the nature of the gain medium. For atomic media it was first recognized by Lax \cite{lax_conf_1966} and tends to be relatively 
small; for semiconductor media its importance was realized by Henry \cite{henry82,henry86}, and in this context it typically 
dominates the direct phase fluctuation terms found by Schawlow and Townes, and is called the Henry $\alpha$ factor.  A second 
correction arises from relaxing the assumption of complete inversion of the gain medium used by Schawlow-Townes; this 
incomplete inversion factor accounts for the actual number of inverted gain atoms  \cite{lax_conf_1966}. A third and particularly interesting correction is the Petermann factor, 
which describes the effect of the openness of the cavity and the consequent non-orthogonality of the lasing modes
\cite{petermann_calculated_1979,haus_excess_1985,siegman_excess_1989,hamel_nonorthogonality_1989,hamel_observation_1990}. 
Each of these corrections {\it increases} the linewidth from the ST value, which it is natural to regard as a lower bound.
However, there exists a fourth correction, often referred to as the ``bad-cavity'' factor, which leads to a {\it reduction} in the laser linewidth.  
This correction is only appreciable when the cavity decay rate, $\gamma_0$ is fast enough (Q is low enough) that 
 $\gamma_0 \sim \gamma_\perp$, where the latter is the dephasing rate of the polarization of the gain medium, which determines the gain 
bandwidth \cite{lax_conf_1966,haken_LT,kolobov_role_1993,kuppens_quantum-limited_1994,vanexter_theory_1995,kuppens_evidence_1995}. Hence it is appreciable when gain dispersion is significant.
This correction was first
interpreted as a slowing of phase diffusion due to atomic memory effects \cite{lax_conf_1966,haken_LT,kolobov_role_1993}, and
subsequently an alternative interpretation was pointed out: the dispersion reduces the group velocity of the light
within the cavity, leading to an increase in the effective cavity Q and a narrowing of the ST linewidth due to the active medium
 \cite{kuppens_quantum-limited_1994}. More recently, superradiant gain media have been proposed as a way of
using the bad-cavity factor to achieve ultralow linewidth lasers \cite{meiser_spin_2008,meiser_prospects_2009,meiser_steady-state_2010,meiser_intensity_2010,bohnet_steady-state_2012}.

However none of the previous linewidth theories have treated fully the space-dependence of the electric fields
and the non-linear spatial hole-burning effect in lasers, which greatly affects the stimulated and spontaneous emission rates
at different points in the cavity.  Recently, a steady-state \textit{ab initio} laser theory (SALT) \cite{tureci06,tureci08,ge10} has 
been developed which treats the spatial degrees of freedom essentially exactly, even in the case of multimode lasing. For single-mode lasing the 
validity of the theory only requires that $\gamma_\perp \gg \gamma_\parallel$,where $\gamma_\parallel$ is the non-radiative relaxation rate of the lasing transition; 
for multimode lasing the condition $\gamma_\parallel \ll \Delta$, where $\Delta$ is the free spectral range of the resonator, is also required \cite{ge08,ge10,esterhazy14}.  Subsequently ab initio
linewidth theories, based on fluctuations in the fields around the SALT solutions, have led to generalized linewidth formulas which should be more accurate than the 
original ST linewidth formula with the four previous corrections included as independent multiplicative factors, as is typically done. 
The first works of this type used a scattering matrix formulation of the quantum fluctuations and input-output theory \cite{chong12,pillay_2014}, 
which captured correctly the generalization of the Petermann and bad-cavity factors, but not that of the alpha and incomplete inversion factors.  The linewidth is 
expressed in terms of the residue of the lasing pole in the scattering matrix and leads to analytic 
formulas in terms of the SALT solutions.  Very recently Pick \textit{et al.} \cite{pick_linewidth_2015} have derived a more general analytic formula for the
linewidth, by applying a coupled mode noise analysis to the SALT solutions.  This formula agrees with the results of \cite{chong12,pillay_2014} but goes beyond 
them to include correctly more general $\alpha$ and incomplete inversions factors.
We will refer to this generalized theory, which includes noise effects, as N-SALT (SALT plus noise).
We believe that the N-SALT linewidth formula quantitatively predicts the laser linewidth (far above threshold) including 
all corrections in an appropriately generalized form, and in that sense represents completely the effects of spontaneous
emission on the laser linewidth.  We test this hypothesis in the current work by direct integration of the laser equations with noise. 

Adding Langevin noise to the steady-state lasing solutions for a gain medium of two-level atoms was shown in \cite{pick_linewidth_2015} to lead to a set of non-linear 
coupled mode equations for the time-dependent fluctuations around the SALT steady-state.  Evaluation of the 
noise-averaged field correlation functions from these equations gives the N-SALT laser linewidth in the form:
\begin{align}
\dwPick =& \frac{\hbar \omega_0}{2 P}\frac{\omega_0^2\int \im[\varepsilon(\mathbf{x},\omega_0)] |\boldsymbol{\psi}_0(\mathbf{x})|^2 d\mathbf{x} \int \im[\varepsilon(\mathbf{x},\omega_0)]\frac{N_2(\mathbf{x})}{D(\mathbf{x})}|\boldsymbol{\psi}_0(\mathbf{x})|^2 d\mathbf{x}}
{\left|\int \boldsymbol{\psi}_0^2(\mathbf{x}) \left(\varepsilon(\mathbf{x},\omega_0) + \frac{\omega_0}{2} \frac{d\varepsilon}{d\omega}|_{\omega_0} \right) d\mathbf{x} \right|^2} (1 + \tilde{\alpha}^2), \label{eq:noiseFDTDCMT}
\end{align}
where $\boldsymbol{\psi}_0(\mathbf{x})$ is the the semiclassical lasing field inside of the cavity found from SALT, normalized such that 
$\int \boldsymbol{\psi}_0^2 d\mathbf{x} = 1$, and the integral is over the cavity region. 
$\varepsilon(\mathbf{x})$ is the total dielectric function of the passive cavity plus gain medium,
assumed here to be homogeneously broadened two-level atoms, and
$N_2(\mathbf{x})$ and $D(\mathbf{x})$ are the number of excited atoms and the atomic inversion respectively (generalization
to multi-level, multi-transition atoms is straightforward within SALT and N-SALT, see \cite{cerjan12,cerjan_csalt_2015}).
$\tilde{\alpha}$ is the generalized $\alpha$ factor \cite{pick_linewidth_2015}, which can be calculated analytically from knowledge of 
$\boldsymbol{\psi}_0(\mathbf{x}),\varepsilon(\mathbf{x})$\cite{pick_linewidth_2015}.  This formula is derived under the conditions that 
$\dwPick \ll \gamma_\parallel,\Delta$. 
This equation reduces to the separable corrections discussed above in the appropriate limits \cite{pillay_2014,pick_linewidth_2015},
but show that in general the incomplete inversion, Petermann, and bad-cavity
linewidth corrections cannot be considered independent from each other or of the cavity
decay rate. 

Here, we test the predictions of the N-SALT linewidth formula against 
the Schawlow-Townes linewidth formula, including all the relevant corrections
by directly integrating the laser equations using
the Finite Difference Time Domain (FDTD) method, including the quantum fluctuations using
the method proposed by Drummond and Raymer \cite{drummond_raymer_1991}, and employing the time-stepping method proposed by Bid\'{e}garay \cite{bidegaray03}. 
Many previous numerical studies of spontaneous emission in laser cavities have implemented
the noise based on knowledge of the lasing mode structure \cite{marcuse_computer-simulation_1984-1,marcuse_computer-simulation_1984,gray_noise_1989,kira_quantum_1999}.
However, these studies did not have access to the above-threshold lasing-mode profiles,
which can differ significantly from the passive cavity modes used e.g. in
calculating the traditional Petermann factor. In our approach we will not make a particular modal ansatz.
Hofmann and Hess derived FDTD-based noisy lasing equations similar to ours for applications
to semiconductors, but the analysis
made further assumptions not valid above the lasing threshold \cite{hofmann_quantum_1999}.
The effects of fluctuations in the electromagnetic fields due to thermal noise
has also been previously studied using the FDTD algorithm \cite{luo_thermal_2004,andreasen_finite-difference_2008,rodriguez_casimir_2009}; these effects 
are necessary to include when studying the noise properties of masers or other long wavelength
lasers, but can be safely neglected at optical frequencies, where the 
spontaneous emission events being considered here dominate the noise of the laser.
The approach used in this manuscript is similar to that used by Andreasen \textit{et al.}\ \cite{andreasen_fdtd_2009,andreasen_numerical_2010,andreasen_thesis},
both in the equations used and in the analytic method to extract the signal's linewidth. However 
unlike those earlier studies \cite{andreasen_fdtd_2009,andreasen_numerical_2010,andreasen_thesis}
we will analyze the linewidth far above threshold where it can be compared 
quantitatively to previous proposed formulas. To our knowledge this is the first study of this type.
To this end, we will be considering relatively simple and small
laser cavities, allowing us to achieve the spectral resolution necessary to resolve
the narrow laser linewidths far above the lasing threshold.

The outline of the remainder of this paper is as follows. In Sec.~\ref{sec:noiseFDTDtwo} we demonstrate
the equivalence of the macroscopic picture of the N-SALT linewidth formula with the microscopic
picture used by Drummond and Raymer. In Sec.~\ref{sec:fdtd} we review the equations
and numerical method used in the FDTD algorithm to simulate a noisy gain medium coupled
to a laser cavity. Sec.~\ref{sec:four} presents the methodologies for extracting a
linewidth from the resultant noisy signal in both the frequency and time domains. The
results of our study are given in Sec.~\ref{sec:results}, including the direct
comparison between the Schawlow-Townes and N-SALT linewidth predictions in a simple laser cavity with
single mode lasing, in single mode lasers with a relatively large $\alpha$ factor, and in the case 
of multimode lasing, particularly near the second lasing threshold.
Summary and concluding remarks are given in Sec.~\ref{sec:summary}.

\section{Microscopic and macroscopic noise equivalence \label{sec:noiseFDTDtwo}}

There are two different ways of incorporating the effects of spontaneous emission
on the electric field inside of the laser cavity, either by using the fluctuation-dissipation
theorem alongside the wave equation, or by including spontaneous emission in the
atomic degrees of freedom, which are coupled non-linearly to the wave equation.
In this section we will explicitly demonstrate the equivalence of these two methods,
which we term the macroscopic and microscopic perspectives respectively, as the
derivation of the N-SALT linewidth equation uses the former method, while the Langevin
equations augmenting the FDTD simulations use the latter. This section also serves as
a proof that despite the non-equilibrium nature of the laser, with power flowing in and
light flowing out, the system does reach a point of stability wherein the fluctuations
of the electric field can be appropriately treated with the fluctuation-dissipation theorem.

The derivation of the N-SALT equation incorporates
all of the noise due to the quantum fluctuations in the gain medium directly
into the wave equation as \cite{pick_linewidth_2015}
\begin{equation}
\left[ \nabla \times \nabla \times - \omega^2 \varepsilon(\omega, \mathbf{E}_0)\right] \mathbf{E} = \omega^2 \left(\varepsilon(\omega, \mathbf{E}) - \varepsilon(\omega, \mathbf{E}_0)\right) \mathbf{E} + \mathbf{F}_S, \label{eq:noiseFDTDsec1e1}
\end{equation}
where $\varepsilon(\omega,\mathbf{E})$ is the full dielectric function of the cavity and gain medium,
$\varepsilon(\omega,\mathbf{E}_0)$ is the non-linear saturated dielectric function of the cavity evaluated using the
semiclassical lasing mode $\mathbf{E}_0(\mathbf{x}) = \sqrt{I}\boldsymbol{\psi}_0(\mathbf{x})$, where $I$ is
the lasing mode intensity, and
$\mathbf{F}_S$ is a random noise source corresponding to the spontaneous
emission from the gain medium. The first term on the right hand side
of Eq.~(\ref{eq:noiseFDTDsec1e1}) corresponds to the effective source due
to fluctuations in the field leading to fluctuations in the saturation of
the gain medium, while the second term corresponds to spontaneous emission
contributing directly to noise in the electric field.  The inclusion of  the full space-dependent 
non-linearity of the active cavity dielectric
function above threshold in the noise term is a key feature 
distinguishing N-SALT from previous linewidth theories.
The autocorrelation of the random noise
source is assumed to be given directly by the fluctuation-dissipation theorem,
\begin{equation}
\langle \mathbf{F}_S^\dagger(\mathbf{x},\omega)\mathbf{F}_S(\mathbf{x}',\omega') \rangle = 2 \hbar \omega^4 \im[\varepsilon(\omega,\mathbf{E}_0)] \coth\left(\frac{\hbar\omega\beta(\mathbf{x})}{2}\right) \delta(\mathbf{x}-\mathbf{x}')
\delta(\omega-\omega'), \label{eq:macroFScorrPRE}
\end{equation}
where $\beta(\mathbf{x}) = (1/\hbar \omega_0) \ln (N_1(\mathbf{x})/N_2(\mathbf{x}))$ is the effective (negative) inverse temperature 
of the inverted gain medium, with $N_1$ and $N_2$ are the number of atoms in the ground and excited
atomic levels respectively. (Note that $\im[\varepsilon(\omega,\mathbf{E}_0)] <0$ in the inverted state, so that the
correlation remains positive).

In this treatment of the noise in the laser field due to spontaneous emission,
the atomic degrees of freedom have been completely integrated out, and the
fluctuation-dissipation theorem has been invoked from a macroscopic perspective,
relating the autocorrelation of the noise source to the imaginary part of the
material response function and a temperature dependent term. 
The hyperbolic cotangent factor arrises as a sum of a Bose-Einstein
distribution and a factor of $1/2$ from the quantum zero-point fluctuations,
which is why the auto-correlation does not vanish in the zero temperature limit ($\beta \to \infty$).
However, it was shown by Henry and Kazarinov that the contributions from the zero-point
fluctuations cancel in the linewidth formula \cite{henry_quantum_1996} (a simpler, semiclassical proof of this
is in Ref.~\cite{pick_linewidth_2015}), and as such it is convenient
to explicitly subtract this contribution, allowing for the effective temperature
of the gain medium to be determined by relative occupations of the atomic levels comprising the lasing transition,
\begin{equation}
\frac{1}{2}\left[\coth\left(\frac{\hbar \omega_0 \beta(\mathbf{x})}{2}\right) - 1\right] = -\frac{N_2(\mathbf{x})}{D(\mathbf{x})}, \label{eq:temp}
\end{equation}
where $D(\mathbf{x}) = N_2(\mathbf{x}) - N_1(\mathbf{x})$ is the number of inverted atoms.
Thus, for the laser systems considered here, Eq.~(\ref{eq:macroFScorrPRE}) can be written as
\begin{equation}
\langle \mathbf{F}_S^\dagger(\mathbf{x},\omega)\mathbf{F}_S(\mathbf{x}',\omega') \rangle = 4 \hbar \omega^4 \im[\varepsilon(\mathbf{x},\omega)] \left[\frac{1}{2}\coth\left(\frac{\hbar \omega_0 \beta(\mathbf{x})}{2}\right) - \frac{1}{2}\right] \delta(\mathbf{x}-\mathbf{x}') \delta(\omega-\omega'). \label{eq:macroFScorr}
\end{equation}

In contrast to this macroscopic picture, many traditional theories of the noise due to spontaneous emission from the gain media begin by treating the 
Langevin forces on the quantum operators of individual gain atoms and building
up an understanding of the total noise this generates in the electric field, a
more microscopic viewpoint \cite{haken,lax_conf_1966,drummond_raymer_1991}. 
We will demonstrate the equivalence of these two methods by deriving the total Langevin force on the polarization from
the microscopic perspective. For a two-level atomic gain medium, 
the evolution equation for the off-diagonal matrix element of the $\alpha$th atom, $\rho_{21}^{(\alpha)}$, including
the Langevin force, $\Gamma_{(\rho)}^{(\alpha)}(t)$, is given by,
\begin{equation}
\partial_t \rho_{21}^{(\alpha)}(t) = -(\gamma_\perp + i \omega_a) \rho_{21}^{(\alpha)}(t) + \frac{i d^{(\alpha)}}{\hbar} \boldsymbol{\theta} \cdot \mathbf{E}(\mathbf{x}^{(\alpha)},t) + \Gamma_{(\rho)}^{(\alpha)}(t), \label{eq:microeq1}
\end{equation}
in which $\omega_a$ is the atomic transition frequency, $\gamma_\perp$ is the dephasing rate,
and $\boldsymbol{\theta}$ is the dipole coupling matrix element.
Furthermore, the evolution of the inversion for that atom, $d^{(\alpha)}$, including the Langevin force, $\Gamma_{(d)}^{(\alpha)}(t)$, is given by
\begin{equation}
\partial_t d^{(\alpha)} = \gamma_\parallel(d_0^{(\alpha)} - d^{(\alpha)}) + \frac{2}{i\hbar} \boldsymbol{\theta} \cdot \mathbf{E}(\mathbf{x}^{(\alpha)},t)(\rho_{21}^{(\alpha)*} - \rho_{21}^{(\alpha)}) + \Gamma_{(d)}^{(\alpha)}(t) \label{eq:microeq2}
\end{equation}
where $d_0^{(\alpha)}$ is the inversion of the $\alpha$th atom in the absence of any electric field.
Finally, the wave equation for the electric field can be written in this context by explicitly
including the coupling between the field and each individual gain atom (see Eqs.~(5.48) and (5.55) in Ref.~\cite{haken}),
\begin{equation}
\left[ \nabla \times \nabla \times -\omega_0^2 \varepsilon_c \right]\mathbf{E}(\mathbf{x},\omega) =  4\pi \omega_0^2 \boldsymbol{\theta} \sum_{\alpha} \delta(\mathbf{x} - \mathbf{x}^{(\alpha)})\rho_{21}^{(\alpha)}, \label{eq:microWave}
\end{equation}
in which we have approximated that the electric field is oscillating at 
frequencies close to the semiclassical lasing frequency, $\omega_0$,
and retained only the positive frequency components for both the electric
field and atomic polarization.
Our aim is to determine the form of the effective total Langevin force on the electric field
by solving Eqs.~(\ref{eq:microeq1}) and (\ref{eq:microeq2}) for the polarization and inversion, 
insert these expressions into the wave equation, and collect the
resulting Langevin force terms.

To leading order, $\rho_{21}$ will
oscillate at the lasing frequency, $\omega_0$, and if we approximate this as its only frequency component,
we can solve for
\begin{equation}
{\rho}_{21}^{(\alpha)} = \frac{-d^{(\alpha)}}{\hbar(\omega_0 - \omega_a + i \gamma_\perp)}\boldsymbol{\theta} \cdot \tilde{\mathbf{E}}(\mathbf{x}^{(\alpha)},\omega) + \frac{i e^{i \omega_0 t}}{\omega_0 - \omega_a + i \gamma_\perp} \Gamma_{(\rho)}^{(\alpha)}, \label{eq:microRho}
\end{equation}
where the electric field is assumed to be a constant over the volume of the atom at $\mathbf{x}^{(\alpha)}$.
The fluctuation dissipation theorem states that the strength of the fluctuations is proportional
to the strength of the dissipative terms. Thus, for the Class A and B lasers considered here,
$\gamma_\parallel \ll \gamma_\perp$, so $\Gamma_{(d)}^{(\alpha)}(t) \ll \Gamma_{(\rho)}^{(\alpha)}(t)$, and
we can safely ignore the fluctuations in the atomic inversion. Thus, we can insert Eq.~(\ref{eq:microRho}) into
Eq.~(\ref{eq:microWave}),
\begin{multline}
\left[ \nabla \times \nabla \times - \omega_0^2 \varepsilon_c \right] \mathbf{E}(\mathbf{x},\omega) = 4 \pi \omega_0^2 \boldsymbol{\theta} \sum_{\alpha} \delta(\mathbf{x} - \mathbf{x}^{(\alpha)}) \\
\left[\frac{-d^{(\alpha)}(\boldsymbol{\theta} \cdot \mathbf{E}(\mathbf{x}^{(\alpha)},\omega)}{\hbar(\omega_0 - \omega_a + i \gamma_\perp)}
 + \frac{i e^{i\omega_0 t}}{\omega_0 - \omega_a + i \gamma_\perp} \Gamma_{(\rho)}^{(\alpha)} \right]. \label{eq:microfullfield}
\end{multline}
Equation (\ref{eq:microfullfield}) allows for the identification of the spontaneous noise in the polarization, $\mathbf{P}_N$,
using Eq.~(\ref{eq:noiseFDTDsec1e1}) and noting that $\mathbf{F}_S = -4 \pi \omega^2 \mathbf{P}_N$,
as
\begin{equation}
\mathbf{P}_N(\mathbf{x},\omega) = \sum_{\alpha} \delta(\mathbf{x}-\mathbf{x}^{(\alpha)}) \frac{i \boldsymbol{\theta}e^{i\omega_0t}}{\omega_0 - \omega_a + i\gamma_\perp} \Gamma_{(\rho)}^{(\alpha)}(\omega).
\end{equation}
We can now directly calculate the correlation function of the spontaneous noise in the polarization using
the correlation of the atomic Langevin force \cite{haken},
\begin{equation}
\langle \Gamma_{(\rho)}^{(\alpha)}(t) \Gamma_{(\rho)}^{(\beta)\dagger}(t') \rangle = \left[\gamma_\perp 
(1 + \langle d^{(\alpha)} \rangle) + \frac{\gamma_\parallel}{2}(d_0^{\alpha} - \langle d^{(\alpha)} \rangle)\right] \delta_{\alpha \beta} \delta(t-t'), \label{eq:noiseFDTDLangGain}
\end{equation}
in which each atom is taken to only be in equilibrium with its reservoir \cite{haken_LT}. Note that the only place
the non-equilibrium nature of the reservoir comes in is via the term $\langle d^{(\alpha)} \rangle = \rho^\alpha_{22}
- \rho^\alpha_{11}$.  Since no higher moments or correlations enter the calculation, it is safe to define an effective
temperature for the system which can be negative via this relation and apply the fluctuation-dissipation theorem.
Note also that $\langle d^{(\alpha)} \rangle$ contains the non-linear effect of gain saturation and spectral hole burning 
when calculated by the FDTD method given below.

By assuming that the inversion is relatively stationary, we can identify the
same frequency auto-correlation of the noise as \cite{LLvol5}
\begin{equation}
\langle \Gamma_{(\rho)}^{(\alpha)}(\omega) \Gamma_{(\rho)}^{(\beta)\dagger}(\omega) \rangle = \gamma_\perp (1 + \langle d^{(\alpha)} \rangle) \delta_{\alpha \beta},
\end{equation}
in which we have again dropped the noise source proportional to $\gamma_\parallel$, to be consistent
with the approximation neglecting fluctuations in the inversion made above. This allows us to solve for
\begin{equation}
\langle \mathbf{P}_N^{\dagger} (\mathbf{x},\omega) \mathbf{P}_N(\mathbf{x}',\omega) \rangle = \frac{2 \boldsymbol{\theta}^2 \gamma_\perp}{(\omega_0 - \omega_a)^2 + \gamma_\perp^2} N_2(\mathbf{x}) 
\delta(\mathbf{x}-\mathbf{x}'), \label{eq:microPNcorr}
\end{equation}
where the number of atoms in the upper lasing state, $N_2(\mathbf{x})$ has been identified using,
\begin{equation}
N_2(\mathbf{x}) = \frac{1}{2}\sum_\alpha \delta(\mathbf{x}-\mathbf{x}^{(\alpha)}) (1 + \langle d^{(\alpha)} \rangle).
\end{equation}
Upon substitution of the imaginary part of the
dielectric function,
\begin{equation}
\im[\varepsilon] = -\frac{4\pi \boldsymbol{\theta}^2}{\hbar} \frac{\gamma_\perp D(\mathbf{x})}{(\omega - \omega_a)^2 + \gamma_\perp^2},
\end{equation}
we can identify
the same frequency auto-correlation of the noise source $\mathbf{F}_S$ as
\begin{equation}
\langle \mathbf{F}_S^{\dagger} (\mathbf{x},\omega) \mathbf{F}_S(\mathbf{x}',\omega) \rangle = 8\pi\omega_0^4 \hbar \im[\varepsilon] \frac{N_2(\mathbf{x})}{D(\mathbf{x})} 
\delta(\mathbf{x}-\mathbf{x}').
\end{equation}
Finally, noting that the different frequency auto-correlation function can be
found as \cite{LLvol5},
\begin{equation}
\langle \mathbf{F}_S^{\dagger} (\mathbf{x},\omega) \mathbf{F}_S(\mathbf{x}',\omega') \rangle = \frac{1}{2\pi} \langle \mathbf{F}_S^{\dagger} (\mathbf{x},\omega) \mathbf{F}_S(\mathbf{x}',\omega) \rangle \delta(\omega - \omega')
\end{equation} 
and using the definition of the temperature factor given in Eq.~(\ref{eq:temp}),
we recover the expected auto-correlation of the random noise source given in Eq.~(\ref{eq:macroFScorr}).
With this, we have verified that the microscopic and macroscopic methods of treating
the fluctuations in the gain medium produce identical results, which allows us to use
a microscopic model of the gain medium in our FDTD simulations to test the predictions of the N-SALT theory.

\section{FDTD equations \label{sec:fdtd}}

Having now demonstrated the equivalence of the microscopic and macroscopic fluctuation models,
in this section we show how to include the microscopic fluctuations of the gain medium
within an FDTD simulation of a laser.
The FDTD algorithm has been known since the 1960s \cite{yee66} and is ubiquitous across
many fields of study \cite{taflove}. However, only a few previous works have used the algorithm to study the noise in lasers
\cite{andreasen_fdtd_2009,andreasen_numerical_2010,andreasen_thesis}, and none (to our knowledge)
have studied the linewidth far above the lasing threshold as we do here. For this reason we will briefly review the 
simulated equations here. The Maxwell-Bloch equations for a two level atomic gain medium in
a one dimensional cavity can be written as
\begin{align}
  \frac{d}{dt} E_n =& \frac{c^2}{\varepsilon_c} \left[ \frac{d}{dx} B_n + 4 \pi \left(\frac{\theta}{V_0} \right) \frac{d}{dt} \left(J_n^- + (J_n^-)^* \right) \right], \\
  \frac{d}{dt} B_n =& \frac{d}{dx} E_n, \\
  \frac{d}{dt} J_n^- =& -(\gamma_\perp + i \omega_a) J_n^- - \frac{\theta}{i \hbar} E_n D_n + F_n^{(J)}, \\
  \frac{d}{dt} D_n =& -\gamma_\parallel(D_n - D_{0,n}) + \frac{2 \theta}{i \hbar}E_n( (J_n^-)^* - J_n^-) + F_n^{(D)},
\end{align}
where $E_n$ and $B_n$ are the electric and magnetic field densities at the spatial location $x_n$ within
the lasing cavity, $V_0$ is the volume associated with each grid point, 
$J_n^-$ is the total atomic off-diagonal density matrix element (related to the polarization) with
a positive frequency component, $D_n$ is the inversion
of the $N_n$ atoms at the spatial location $x_n$, $D_{0,n}$ is the
inversion in the absence of an electric field and plays the role of the effective pump strength in this theory, and
$F_n^{(J)}$ and $F_n^{(D)}$ are the Langevin forces experienced by the atomic
off-diagonal density matrix element and inversion respectively. The choice of $J_n$ for the total off-diagonal density
matrix element is made for ease of comparison
with Drummond and Raymer, who use $J_n^-$ to denote the same quantity, and is defined as
\begin{equation}
J_n(x) = \sum_\alpha \rho_{21}^{(\alpha)} \delta(x-x^{(\alpha)}) = N_n \rho_{21}(x).
\end{equation} 
The Langevin forces can be written as \cite{drummond_raymer_1991},
\begin{align}
  F_n^{(J)} =& \xi_n^{(J)} \sqrt{ - 2 i \theta E_n J_n^-} + \xi_n^{(P)} \sqrt{\gamma_P(D_n + N_n)} + \xi_n^{(N)}\sqrt{\gamma_{21,n} N_n}, \label{F1} \\
  F_n^{(D)} =& 2\xi_n^{(D)} \left[ \frac{\gamma_\parallel}{2} (N_n - \frac{D_{0,n}}{N_n} D_n) + i \theta(J_n^- E_n - J_n^+ E_n) - 2\gamma_{21,n} \frac{J_n^+ J_n^-}{N_n} \right]^{(1/2)} \notag \\
  &- 2\left[\xi_n^{(N)} J_n^+ + \xi_n^{(N)*} J_n^- \right] \sqrt{\frac{\gamma_{21,n}}{N_n}}, \label{F2}
\end{align}
in which $\gamma_{21}$ is the pumping rate from lower level $|1\rangle$ to $|2 \rangle$ and
is given by,
\begin{equation}
  \gamma_{21,n} = \frac{\gamma_\parallel}{2}\left(1+ \frac{D_{0,n}}{N_n} \right),
\end{equation}
and $\gamma_P = \gamma_\perp - \gamma_\parallel/2$ is the pure dephasing rate. Randomness is introduced
through the stochastic variables $\xi$, which are complex except for $\xi_n^{(d)} \in \mathbb{R}$, 
and satisfy \cite{drummond_raymer_1991}
\begin{equation}
\langle \xi_n^{(i)}(t) \xi_m^{(j)}(t') \rangle = \delta(t - t') \delta_{nm} \delta_{ij}. \label{eq:XIdef}
\end{equation}
Many of the terms in Eqs.~(\ref{F1}--\ref{F2}) stem from resolving the dilemma of the operator
ordering when reducing operator equations to c-number equations. However, for 
studying the laser linewidth above threshold, the difference caused by this
ambiguity is minimal, as the addition or removal of a spontaneous emission event
is negligible in the presence of the large number of gain atoms necessary for lasing to occur.
Thus most of these terms are expected to be negligible, an assumption which we check \textit{a posteriori} after retaining
the leading terms:
\begin{align}
  F_n^{(J)} =& \xi_n^{(P)} \sqrt{\gamma_P(D_n + N_n)} + \xi_n^{(N)}\sqrt{\gamma_{21,n} N_n}, \label{F3} \\ 
  F_n^{(D)} =& 2\xi_n^{(D)} \sqrt{ \frac{\gamma_\parallel}{2} \left(N_n - \frac{D_{0,n}}{N_n} D_n \right)}. \label{F4}
\end{align}
Finally, in accordance with the discussion in the previous section,
the thermal fluctuations of the electric and magnetic fields have been neglected.

The Maxwell-Bloch equations can then be discretized for use in the FDTD algorithm following
the weak coupling method proposed by Bid\'{e}garay\cite{bidegaray03}, evolving the atomic variables
simultaneously with the magnetic field, but at the same spatial locations as the electric field
so as to avoid solving a non-linear equation. Furthermore, it is useful to separate the
real and imaginary components of the atomic off-diagonal density matrix element, $J_n^- = j_n^{(1)} + i j_n^{(2)}$,
resulting in
\begin{align}
  E_n(t_{i+1}) =& E_n(t_i) + \frac{c^2 \Delta t}{\varepsilon_c} \left[  
    8 \pi \left(\frac{\theta}{V_0} \right) \left(\omega_a j_n^{(2)}(t_{i+\Half}) - \gamma_\perp j_n^{(1)}(t_{i+\Half}) \right) \right. \notag \\
    & \left. + \frac{B_{n+\Half}(t_{i+\Half}) - B_{n-\Half}(t_{i+\Half})}{\Delta x} \right], \label{fdtd1}\\
  B_{n+\Half}(t_{i+\Half}) =& B_{n+\Half}(t_{i-\Half}) +\frac{\Delta t}{\Delta x} \left(E_{n+1}(t_i) - E_{n}(t_i)\right), \\
  \mathbf{u}_n(t_{i+\Half}) =& \left(\frac{1}{\Delta t} I - \frac{1}{2} M\right)^{-1}\left[ \mathbf{d}_n + \mathbf{f}_n + \left(\frac{1}{\Delta t} I + \frac{1}{2} M\right) \mathbf{u}_n(t_{i-\Half})\right], \label{fdtd3}
\end{align}
where $\mathbf{u_n} = (D_n, j_n^{(1)}, j_n^{(2)})$ is the vector of the atomic variables, $\mathbf{d}_n = (\gamma_\parallel D_{0,n}, 0, 0)$
is the pumping vector, $I$ is the $3$x$3$ identity matrix, $M$ is a matrix which contains the coupling information between
the atomic variables,
\begin{equation}
M = \left(
\begin{array}{ccc}
-\gamma_\parallel & 0 & -\frac{4\theta}{\hbar} E_n(t_i) \\
0 & -\gamma_\perp & \omega_a \\
\frac{\theta}{\hbar}E_n(t_i) & -\omega_a & -\gamma_\perp \end{array}
\right),
\end{equation}
and $\mathbf{f}_n$ is the Langevin force vector, whose
elements are
\begin{align}
  f_{n,1} =& 2\xi_n^{(1)} \sqrt{ \frac{\gamma_\parallel}{2} (N_n - \frac{D_{0,n}}{N_n} D_n(t_{i-\Half}))}, \\
  f_{n,2} =& \xi_n^{(2)} \sqrt{\gamma_P(D_n(t_{i-\Half}) + N_n)} + \xi_n^{(3)}\sqrt{\gamma_{21,n} N_n}, \\
  f_{n,3} =& \xi_n^{(4)} \sqrt{\gamma_P(D_n(t_{i-\Half}) + N_n)} + \xi_n^{(5)}\sqrt{\gamma_{21,n} N_n}. \label{fdtd6}
\end{align}
where we have renumbered the random variables $\xi_n^{(i)}$, which continue to satisfy Eq.~(\ref{eq:XIdef}),
but are now real, rather than complex, and introduced a factor of $2^{-1/2}$ in this conversion process (except for $\xi_n^{(1)}$, which was real to begin with).
Here we have used the final approximation that the Langevin force vector only depends upon
the inversion at the previous time step, rather than the average of the previous and current
time steps which would result in a non-linear equation \cite{andreasen_thesis}. This is justified
for the simulations performed here because the inversion, $d_n$, is many orders of magnitude smaller
than the total number of atoms, $N_n$, and thus these inversion dependent terms will have minimal impact upon the overall
strength of the noise. For the discretized Langevin forces, the stochastic variables
$\xi_n^{(k)}$ are chosen from a standard uniform distribution, and then renormalized to
satisfy
\begin{equation}
  \langle \xi_n^{(k)}(t_i) \xi_m^{(l)}(t_j) \rangle = \frac{1}{\Delta t} \delta_{ij} \delta_{nm} \delta_{kl}.
\end{equation}
Eqs.~\ref{fdtd1}-\ref{fdtd6} can now be readily evaluated numerically. 

\section{Linewidth analysis \label{sec:four}}

Broadly speaking there are two main ways of extracting a linewidth from a noisy signal;
by either fitting a curve to the frequency domain data or calculating
the cross-correlation of the time domain data \cite{proakis}. Here we will use
both methods an compare them; first we calculate a linewidth from the spectral data and then confirm
this linewidth by calculating $\langle \phi(t') \phi(t) \rangle$, where $\phi(t)$ is the
phase of the electric field.

\subsection{Frequency-domain analysis \label{sec:analysis}}

To analyze the spectrum of the electric field output from the cavity, $E(\omega)$, and find a linewidth, 
we will use the method proposed
by Andreasen \textit{et al.}\ \cite{andreasen_thesis}, and fit the spectrum to a Lorentzian through the use of
an error function. We assume that the noise is Lorentzian,
\begin{equation}
L(\omega) = \left(\frac{2 A}{\pi}\right) \frac{s^2}{(\omega - \omega_0)^2 + s^2}
\end{equation}
where $s$ is the half-width half-maximum of the noise, $\dwFDTD = 2s$. The Lorentz
error function can then be defined as
\begin{equation}
L_{EF}(\omega) = \int_{\omega_0}^\omega L(\omega') d\omega' = \left(\frac{2 A s}{\pi}\right) \arctan \left(\frac{\omega - \omega_0}{s} \right). \label{LEF}
\end{equation}
As such, this integration can be carried out numerically directly upon $E(\omega)$,
and then fit to Eq.~(\ref{LEF}). For all of the data shown in this paper the curve fitting
is carried out using iterative least squares estimation. Performing this integration requires knowledge
of the lasing frequency, $\omega_0$, which is known from the semiclassical SALT calculation.
However, the presence of noise results in a slight shift of the semiclassical lasing frequency \cite{henry82}, and 
the slightly different discretization schemes used between the SALT and FDTD calculations yield an additional shift in the lasing frequency, 
which together lead to a slightly shifted integrated spectrum, both horizontally
and vertically. As such it is useful to include two other fitting
parameters in the Lorentz error function,
\begin{equation}
L_{EF}'(\omega) = \left(\frac{2 A s}{\pi}\right) \arctan \left(\frac{\omega - \omega_0 + d}{s} \right) +c,
\end{equation}
where $d$ plays the role of the horizontal offset and $c$ is the vertical offset. Using this correction,
the calculated linewidths are robust to the choice of $\omega_0$ so long as
the curve fitting algorithm converges.

\begin{figure}
\centering
\includegraphics[width=0.95\textwidth]{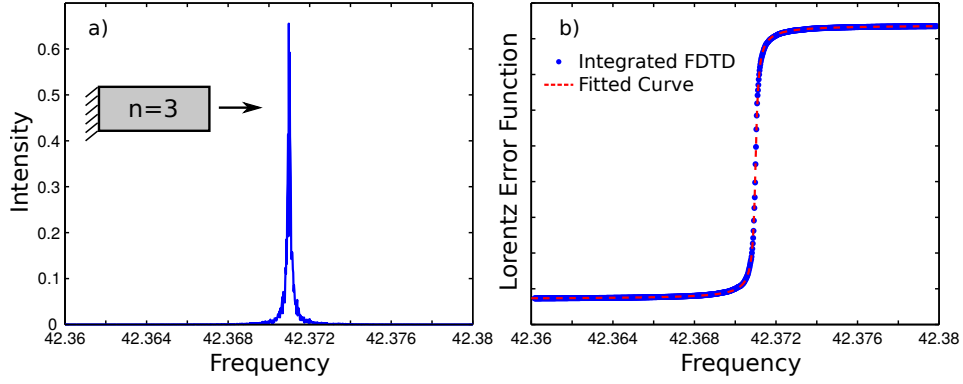}
\caption{(a) Intensity spectrum of the output electric field of an $n=3$ dielectric slab cavity, shown in the schematic. 
The simulation parameters for the cavity are $\gamma_\perp = .5$, 
$\omega_a = 42.4$, $\gamma_\parallel = .01$, $\theta = 2 \times 10^{-9}$, $N_A = 10^{10}$, and the cavity is uniformly pumped at $D_0 = 0.275$ which
is close to $5$ times the threshold lasing pump of $D_{0,thr} = 0.0488$. The rates quoted here are given in units of
$c/L$, while the intensity is given in SALT units of $4 \theta^2 / (\hbar^2 \gamma_\perp \gamma_\parallel)$, and
the number and inversion of gain atoms are given in the SALT units of $4 \pi \theta^2 / (\hbar \gamma_\perp)$. (b) Plot of the
fitted Lorentz error function (red line) and numerically integrated FDTD data (blue dots) of the simulation shown in (a). 
The spectral resolution for the simulated data in (a) and (b) is $d\omega = 1.96 \times 10^{-5}$. The 
analytic curve fit parameters are found using MATLAB's curve fitting algorithms. \label{fig:freqEx}}
\end{figure}

An example of this process can be seen in Fig.~\ref{fig:freqEx}, where the left panel shows
the spectrum of the output electric field for a dielectric slab cavity. To compute the power spectrum, 
or technically the periodogram \cite{proakis} of the noisy signal, we chop the simulated time-domain
field $E(t)$ into $\sim 10$ pieces and perform a discrete-time Fourier transform (DTFT) \cite{fftw} on
each constituent piece, and then ensemble-average the resulting spectra $|\hat{E}(\omega)|^2$ using
Bartlett's method \cite{proakis}. The right panel shows the Lorentz error
function integral calculated numerically and fit against the analytic curve. The resulting
linewidth predicted by this method is $\dwFDTD = 2.22 \times 10^{-4}$, which is around
an order of magnitude larger than the resolution of the resultant spectra,
$d\omega = 1.96 \times 10^{-5}$, given in units of $c/L$.

\subsection{Time-domain confirmation}

This calculation can be independently confirmed by calculating the autocorrelation
of the output electric field as a function of time and expressing this as a function
of the phase correlation, which is defined in terms of the linewidth of the signal.
Writing the output electric field as
\begin{equation}
E(t) = C \cos(\omega t + \phi(t)),
\end{equation}
The autocorrelation of the electric field, $R_{EE}(\delta t) = \langle E(t + \delta t) E(t) \rangle$, can then be written as
\begin{equation}
R_{EE}(\delta t) = \langle E^2(t) \cos(\omega \delta t + \delta \phi(\delta t)) \rangle
- \langle \frac{C^2}{2} \sin(2\omega t + 2 \phi(t)) \sin(\omega \delta t + \delta \phi(\delta t)) \rangle,
\end{equation}
where the double angle formula has been used in finding the second term on the right hand side, and
$\delta \phi(\delta t) = \phi(t + \delta t) - \phi(t)$.
By assuming that the phase shift $\delta \phi(\delta t)$ is uncorrelated with the phase $\phi(t)$,
we can separate the correlations, note that the second term averages to zero, and
again apply a trigonometric identity, resulting in
\begin{equation}
R_{EE}(\delta t) = \frac{C^2}{2} \left[\cos(\omega \delta t) \langle \cos(\delta \phi(\delta t)) \rangle
- \sin(\omega \delta t) \langle \sin(\delta \phi(\delta t))\rangle \right]. \label{eq:xcorr1}
\end{equation}
This assumption that the phase shift, $\delta \phi$ is uncorrelated with the instantaneous
phase, $\phi$, is analogous to assuming that the gain medium has no memory effect,
and is consistent with the earlier assumption that the bad-cavity factor
is unity for the systems studied here. The second term in Eq.~(\ref{eq:xcorr1}) averages
to zero as well, as the phase shift is equally likely to be positive or negative.
Finally, the cosine of the phase shift can be Taylor expanded, and noting the
definition of the linewidth, 
\begin{equation}
\langle \delta \phi^2(\delta t) \rangle = \delta \omega \delta t,
\end{equation}
the electric field autocorrelation can be written as
\begin{equation}
R_{EE}(\delta t) = \frac{C^2}{2}\cos(\omega \delta t) \left[1 - \frac{\delta \omega \delta t}{2} + O(\delta t^2) \right], \label{xcorr}
\end{equation}
showing that in the presence of phase diffusion, the correlation should decrease
linearly for small $\delta t$.

\begin{figure}
\centering
\includegraphics[width=0.6\textwidth]{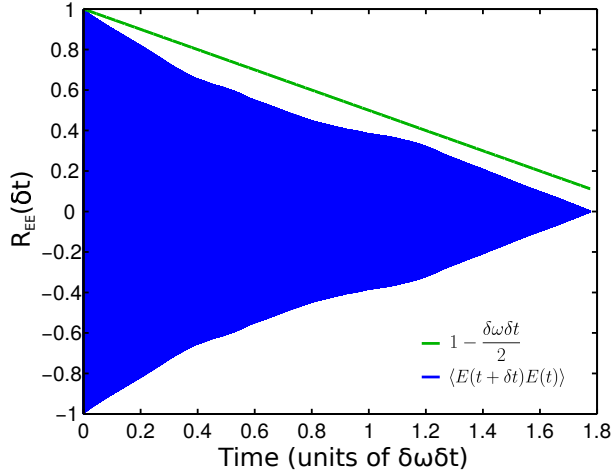}
\caption{Plot of the autocorrelation of the electric field 
simulated numerically for the same parameter used in Fig.~\ref{fig:freqEx} (blue line)
and the analytic prediction for the envelope of the autocorrelation given in the second
factor in Eq.~(\ref{xcorr}) (green line). The fast oscillations in 
the numerically simulated electric field are at the lasing frequency $\omega_0$,
which is much faster than the other time scales in the problem and leads to the
densely packed curve shown in blue.
Quantities are normalized, and plotted in units of
$\delta \omega \delta t$. \label{fig:xcorr}}
\end{figure}

This trend can be observed in Fig.~\ref{fig:xcorr} for the same simulation as shown in
Fig.~\ref{fig:freqEx}, where the prediction for $R_{EE}(\delta t)$ is evaluated
using $\delta \omega$ found by the frequency domain
method from the previous section and Eq.~(\ref{xcorr}) (green line), and numerically
calculated (blue line). The fast oscillations seen in the numerical data are due to
$\omega \delta t \gg 1$, and are predicted by the theory derived above. The semi-quantitative
agreement seen between the frequency domain linewidth prediction and the time domain
prediction calculated here provides a consistency check, though we will use the
frequency domain method for the remainder of the calculations performed here.

\section{Results \label{sec:results}}

To test the predictions of the N-SALT linewidth, given Eq.~(\ref{eq:noiseFDTDCMT}), with the 
Schawlow-Townes linewidth \cite{schawlow_1958}, we first study the simple one-dimensional, single-sided dielectric slab cavity, $n=3$, used in the
previous two sections in Figs.~\ref{fig:freqEx} and \ref{fig:xcorr}. Here, we use the
``fully-corrected'' form of the Schawlow-Townes linewidth as the point of comparison,
which includes the Petermann factor, bad-cavity correction, and Henry $\alpha$ factor, and is given by,
\begin{align}
\dwST^{(corr)} =& \frac{\hbar \omega_0 \gamma_c^2}{2 P}\left(\frac{\bar N_2}{\bar D}\right) \left|\frac{\int |\boldsymbol{\phi}_0(\mathbf{x})|^2 d\mathbf{x}}{\int \boldsymbol{\phi}_0^2(\mathbf{x}) d\mathbf{x}} \right|^2 \left|\frac{1}{1 + \frac{\omega_0}{2\varepsilon}\frac{\partial \varepsilon}{\partial \omega}|_{\omega_0}} \right|^2 (1 + \alpha^2), \label{STcorr}
\end{align}
where $\boldsymbol{\phi}_0(\mathbf{x})$ is the passive cavity resonance corresponding to the lasing mode,
the spatial average of the inversion and occupation of the upper lasing state is denoted as $\bar D = \int D(\mathbf{x}) d\mathbf{x}$,
the spatially averaged inversion is used to calculate the bad-cavity factor, and $\alpha$ is the Henry $\alpha$ factor.
The first term in parentheses of Eq.~(\ref{STcorr}) corresponds to the cavity-averaged incomplete inversion 
factor and the second corresponds to the Petermann factor \cite{petermann_calculated_1979,pillay_2014}.
The quantities $\boldsymbol{\psi}_0(\mathbf{x})$, $\boldsymbol{\phi}_0(\mathbf{x})$, $D(\mathbf{x})$, and $\varepsilon(\mathbf{x})$ are calculated using
SALT, while the FDTD linewidths are extracted using the method described
in Sec.~\ref{sec:analysis}, and run for enough time steps to average together at least six
resulting spectra using Bartlett's method. For the chosen parameters $\gamma_c \sim \gamma_\parallel$
placing it on the border between Class A and Class B lasers \cite{ohtsubo}, close enough to the former
that no relaxation oscillation side-peaks are seen in the resulting spectra.

\begin{figure}
\centering
\includegraphics[width=0.99\textwidth]{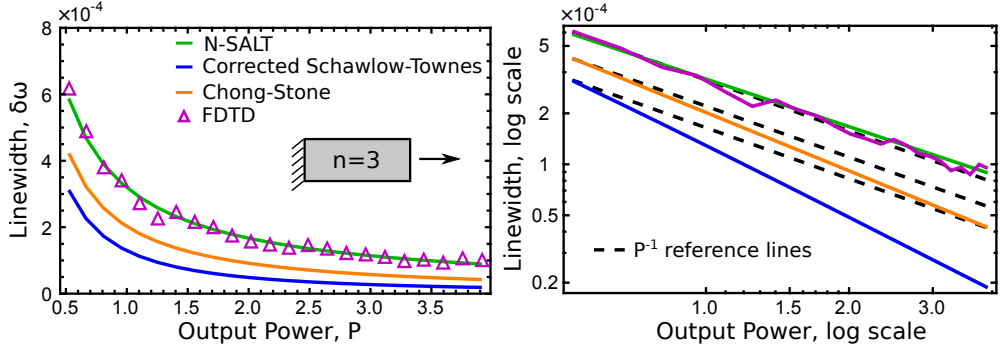}
\caption{(Left panel) Plot showing the linewidth predictions given by the N-SALT given in Eq.~(\ref{eq:noiseFDTDCMT}) (green), corrected
Schawlow-Townes theory given in Eq.~(\ref{STcorr}) (blue), integral form of the Chong-Stone linewidth
formula given in Eq.~(\ref{CSlw}) (orange), and FDTD simulations (magenta) for a uniformly pumped, dielectric slab cavity with $n=3$, $\omega_a = 42.4$,
$\gamma_\perp = .5$, $\gamma_\parallel = .01$, $\theta = 2 \times 10^{-9}$, and $N_A = 10^{10}$.
All of the linewidth formulas are evaluated using the spatially dependent
integral definition of the power given by Eq.~(\ref{goodPower}). (Right panel) Plot of the
same data shown on a log-log scale, with reference lines for strict inverse power dependence, $P^{-1}$,
provided for comparison (black dashed).
Schematic inset shows the cavity geometry.
The rates and frequency are given in units of $c/L$, the number of atoms in the cavity is given 
in terms of the SALT units of $4 \pi \theta^2 / (\hbar \gamma_\perp)$, and the output power is given
in the SALT units of $4 \theta^2 / (\hbar^2 \gamma_\perp \gamma_\parallel)$.
\label{fig:simple}}
\end{figure}

As can be seen in the left panel of Fig.~\ref{fig:simple}, excellent quantitative
agreement is seen between the N-SALT prediction (green line) and the linewidths measured through
direct integration of the noisy Maxwell-Bloch equations (magenta triangles), while both
results differ from the corrected Schawlow-Townes theory (blue line). This discrepancy
is shown to be more than a simple scaling factor in the right panel of Fig.~\ref{fig:simple},
where the same data is plotted on a log-log scale, and it can be seen that the power
law narrowing of the linewidth with respect to the output power differs between the
N-SALT and corrected Schawlow-Townes linewidth predictions. Somewhat surprisingly only
the N-SALT and FDTD results are very close to $P^{-1}$ (black dashed lines), the others are show a measurably
faster narrowing.

To understand the source
of this discrepancy, we also plot the Chong-Stone linewidth \cite{chong12} calculated
using its integral form \cite{pillay_2014},
\begin{equation}
\dwCS = \frac{\hbar \omega_0}{2 P} \left(\frac{\bar{N}_2}{\bar D}\right)\frac{ \left(\omega_0\int \im[\varepsilon(\mathbf{x},\omega_0)] |\boldsymbol{\psi}_0(\mathbf{x})|^2 d\mathbf{x} \right)^2}
{\left|\int \boldsymbol{\psi}_0^2(\mathbf{x}) \left(\varepsilon + \frac{\omega_0}{2} \frac{d\varepsilon}{d\omega}|_{\omega_0} \right) d\mathbf{x} \right|^2} (1 + \alpha^2), \label{CSlw}
\end{equation}
where we have neglected the vanishingly small boundary term (see \cite{pillay_2014}). The Chong-Stone linewidth
formula is derived through considering the behavior of the SALT-based scattering matrix of the cavity,
and thus is able to account correctly for all effects stemming from the cavity; it gives 
the proper cavity decay rate above threshold, and the same Petermann factor,
and bad-cavity correction as N-SALT. However, it does not provide an accurate treatment of the fluctuations
inside the gain medium, particularly amplitude fluctuations, 
and thus is unable to find the $\alpha$ factor and finds an inaccurate, cavity-averaged incomplete inversion factor
similar to conventional theories. For the dielectric slab cavity studied here, the detuning of the lasing
mode from the atomic transition is very small, such that $\alpha \ll 1$. Thus the significant discrepancy 
between the N-SALT and FDTD results and the Chong-Stone prediction
indicates that the largest source of discrepancy lies in the treatment of the incomplete inversion
factor. The ratio of the N-SALT and Chong-Stone linewidth predictions in the limit that $\tilde \alpha = \alpha = 0$ can be written as
\begin{equation}
\frac{\dwCS}{\dwPick} = \frac{\frac{\bar N_2}{\bar D} \int D(\mathbf{x}) |\boldsymbol{\psi}_0(\mathbf{x})|^2 d\mathbf{x}}{\int N_2(\mathbf{x}) |\boldsymbol{\psi}_0(\mathbf{x})|^2 d\mathbf{x}}.
\end{equation}
However, for the two-level atomic gain media simulated here, the number of atoms in the excited
atomic level is nearly constant $N_2 \approx N_1 \approx N/2$, allowing for this ratio to be expressed as
\begin{equation}
\frac{\dwCS}{\dwPick} = \frac{\int D(\mathbf{x}) |\boldsymbol{\psi}_0(\mathbf{x})|^2 d\mathbf{x}}{\int |\boldsymbol{\psi}_0(\mathbf{x})|^2 d\mathbf{x} \int D(\mathbf{x}) d\mathbf{x}}. \label{eq:lwratio}
\end{equation}
In absolute terms, the fluctuations in $N_2$, $N_1$, and $D$ are all of the same magnitude,
but as $D(x) = N_2(x) - N_1(x) \ll N_2$, its spatial variation is much larger on a relative scale and cannot be neglected, 
leading to a significant discrepancy
between the N-SALT/FDTD and Chong-Stone linewidth predictions.  
Note that the approximation of spatial invariance of the occupation of the upper lasing level
does not necessarily hold when considering more realistic gain media, with more than two levels, and is a result of the
well known difficulty in pumping a two-level medium past the transparency point to achieve lasing.  
However the residual discrepancy between Chong-Stone and the corrected ST prediction indicates that
the incomplete inversion factor only accounts for roughly half the discrepancy, and the remainder
(Petermann and bad-cavity effects) would be present in lasers with more than two levels.

\begin{figure}
\centering
\includegraphics[width=0.95\textwidth]{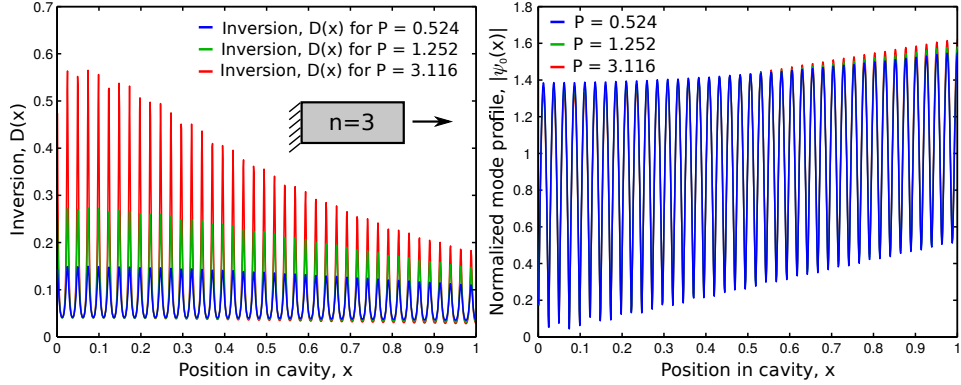}
\caption{(Left panel) Plot of the steady-state inversion, $D(x)$, as a function of the location
in the cavity for three different values of the output power, $P = 0.524$ (blue), $P = 1.252$ (green),
and $P = 3.116$ (red). These values correspond to the first, sixth, and eighteenth data points shown in Fig.~\ref{fig:simple}. 
Strong spatial hole-burning is seen in the inversion due to the lasing mode. Schematic depicts the cavity from Fig.~\ref{fig:simple}.
(Right panel) Plot of the normalized spatial profile of the lasing mode, $|\psi_0(x)|$, as a function
of position in the cavity for the same three values of the output power shown in the left panel. The output power is
given in dimensionless SALT units of $4 \theta^2 / (\hbar^2 \gamma_\perp \gamma_\parallel)$.
\label{fig:invPsi}}
\end{figure}

The implications of the relation expressed in Eq.~(\ref{eq:lwratio}) can be understood graphically
from Fig.~\ref{fig:invPsi}, where the left panel shows the steady-state inversion, $D(\mathbf{x})$, within the cavity for different
values of the output power generated by the cavity, and the right panel shows the spatial dependence of
the lasing mode profile, $|\boldsymbol{\psi}_0(\mathbf{x})|$, for the same values of the output power. As the pump on the
gain medium, $D_0$, is increased, the amplitude of the field within the cavity increases, as does
the output power. However, due to spatial hole-burning in the gain medium, the impact of the higher
field intensity within the cavity is not felt uniformly in the inversion; thus the average inversion
within the cavity still increases as the pump is ramped, mostly due to the positions near the mirror
in the cavity where the electric field is very weak, while the weighted average of the inversion
with the field intensity remains relatively constant, as the inversion where the field intensity is maximized 
stays relatively constant as the pump is increased. Thus as noted, we do expect to see the corrected Schawlow-Townes
and Chong-Stone linewidth predictions decrease faster than $1/P$, as is observed in the right panel
of Fig.~\ref{fig:simple}, as both the output power, $P$, \textit{and} spatially averaged inversion, $\bar D$,
increase as the pump strength, $D_0$, is increased (see Eq.~(\ref{CSlw})). In contrast, the integral of the inversion weighted
against the field intensity, used in the N-SALT linewidth prediction, does not change as the pump is increased.
Thus, even for the two-level atomic gain medium studied here, N-SALT gives a good $1/P$ line narrowing.
Siegman has previously suggested that the incomplete inversion factor might lead to deviations from 
the strict inverse dependence of the laser linewidth upon the output power, but was unable
to test this hypothesis \cite{siegman_observation_1971}.

We note that it is important in these comparisons to calculate the output power from its
fundamental definition via Poynting's theorem \cite{jackson},
\begin{equation}
P = \frac{\omega_0}{2 \pi} \int \im[-\varepsilon(\mathbf{x})] |\mathbf{E}_0(\mathbf{x})|^2 d\mathbf{x}, \label{goodPower}
\end{equation}
where this equation is given in Gaussian units, 
$\mathbf{E}_0(\mathbf{x}) = \sqrt{I} \boldsymbol{\psi}_0(\mathbf{x})$ is the unnormalized lasing mode, and $I$ is
the mode intensity. Performing this calculation relies on finding the correct space-dependent
quantities, which can be obtained using SALT. The quantitative
agreement seen between the N-SALT linewidth prediction and the FDTD simulations
shown in Fig.~\ref{fig:simple} provides independent confirmation that this
is the correct formulation of the output power to use. However, in many treatments of laser 
emisson, which do not treat the full space dependence, the output power is calculated using \cite{haken_LT}
\begin{equation}
P_{\textrm{ST}} = \gamma_c \bar n \hbar \omega_0, \label{badPower}
\end{equation}
where $\bar n$ is the average number of photons in the cavity.  Using this form of power calculation
can introduce a substantial error; using the corrected Schawlow-Townes theory with this spatially 
averaged power for the parameters of Fig.~\ref{fig:simple},  leads to a linewidth roughly a factor of
two larger than the N-SALT and FDTD results.  Thus we see that it is critical to use all of the
spatial information in the fields $\mathbf{E}_0(\mathbf{x})$ and $D(\mathbf{x})$ obtained from SALT in order to
quantitatively predict the laser linewidth.

\subsection{linewidth scaling relations}

The overall intensity of the electric field enters directly
into the linewidth formulas only through the output power, Eq.~(\ref{goodPower}).
SALT demonstrates that the electric field can be written in terms of dimensionless units,
$\mathbf{E}_0(\mathbf{x}) = (\hbar\sqrt{\gamma_\perp \gamma_\parallel} / 2 \theta)\mathbf{E}_{\textrm{SALT}}(\mathbf{x})$ \cite{ge10,li_thesis},
and thus the output power can also be written as,
\begin{equation}
P = \left(\frac{\hbar^2 \gamma_\perp \gamma_\parallel}{4 \theta^2} \right)\frac{\omega_0}{2 \pi} \int \im[-\varepsilon(\mathbf{x})] |\mathbf{E}_{\textrm{SALT}}(\mathbf{x})|^2 d\mathbf{x}.
\end{equation}
This is how the dimension-full parameters stemming
from the properties of the gain medium directly enter into all of the linewidth
formulas discussed here. In particular we can rewrite the N-SALT linewidth in SALT units as,
\begin{equation}
\dwPick = \left(\frac{4 \theta^2}{\hbar^2 \gamma_\perp \gamma_\parallel} \right)
\frac{\hbar \omega_0}{2 P_{\textrm{SALT}}}\frac{\omega_0^2\int \im[\varepsilon] |\boldsymbol{\psi}_0|^2 d\mathbf{x} \int \im[\varepsilon]\frac{N_2}{D}|\boldsymbol{\psi}_0|^2 d\mathbf{x}}
{\left|\int \boldsymbol{\psi}_0^2 \left(\varepsilon + \frac{\omega_0}{2} \frac{d\varepsilon}{d\omega}|_{\omega_0} \right) d\mathbf{x} \right|^2} (1 + \tilde{\alpha}^2), \label{CMTlw}
\end{equation}
where $P_{\textrm{SALT}}$ is the output power calculated using the electric field
measured in SALT units.

\begin{figure}
\centering
\includegraphics[width=0.99\textwidth]{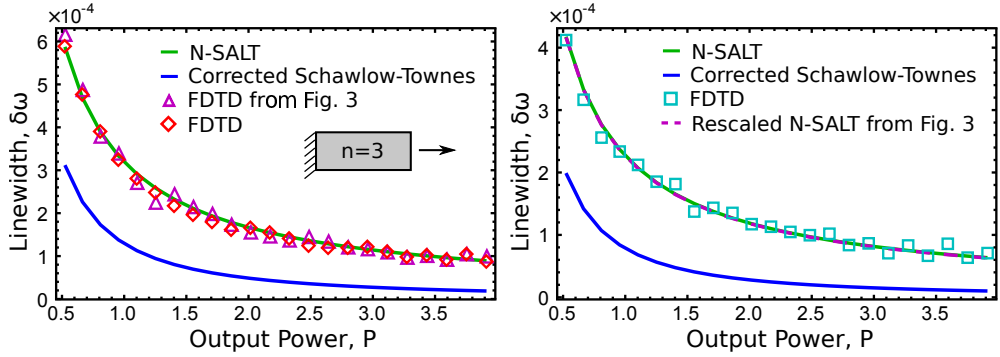}
\caption{(Left panel) Plot showing the linewidth predictions given by the N-SALT (green line), corrected
Schawlow-Townes theory (blue line), and FDTD simulations (red diamonds and magenta triangles) for a uniformly pumped, dielectric slab cavity with $n=3$, $\omega_a = 42.4$,
$\gamma_\perp = .5$, $\gamma_\parallel = .04$, $\theta = 4 \times 10^{-9}$, and $N_A = 10^{10}$, as shown in the schematic.
The results of the new FDTD simulations are shown as red diamonds, and are plotted alongside
the FDTD results from Fig.~\ref{fig:simple}, shown as magenta triangles.
(Right panel) Plot showing the linewidth predictions given by the N-SALT (green line), rescaled N-SALT prediction from Fig.~\ref{fig:simple} (magenta dashed line), corrected
Schawlow-Townes theory (blue line), and FDTD simulations (cyan squares) for a uniformly pumped, dielectric slab cavity with $n=3$, $\omega_a = 42.4$,
$\gamma_\perp = .25$, $\gamma_\parallel = .02$, $\theta = 2 \times 10^{-9}$, and $N_A = 10^{10}$.
The rates and frequency are given in units of $c/L$, the number of atoms in the cavity is given 
in terms of the SALT units of $4 \pi \theta^2 / (\hbar \gamma_\perp)$, and the output power is given
in the SALT units of $4 \theta^2 / (\hbar^2 \gamma_\perp \gamma_\parallel)$.
\label{fig:scale}}
\end{figure}

Using SALT units and the stationary inversion approximation implies
powerful scaling relations between lasing solutions at different gain medium parameter values \cite{tureci06,ge10}.
Similarly, Eq.~(\ref{CMTlw}) implies various scaling relations for the linewidth.
It separates out the dependence of the intrinsic laser linewidth upon
$\theta$, $\gamma_\parallel$, and the leading dependence upon $\gamma_\perp$ and thus
predicts that the linewidth should obey a set of scaling relations.  For example maintaining the ratio
of $\gamma_\parallel / \theta^2$ should yield the same linewidth, and keeping
the ratio $ \gamma_\perp \gamma_\parallel / \theta^2$ constant should result in
only very modest changes in the linewidth (changing $\gamma_\perp$ only changes
the strength of the bad-cavity correction). These predictions are confirmed by FDTD
simulations. In the left panel of Fig.~\ref{fig:scale}, the linewidth is calculated via FDTD  for a
different value of $\gamma_\parallel$ and $\theta$ than in Fig. 3, while keeping the ratio $\gamma_\parallel / \theta^2$ 
to that in Fig.~\ref{fig:simple}.   The resulting
FDTD linewidth (red diamonds, plotted alongside magenta triangles from Fig.~\ref{fig:simple}) is seen to be identical.
This serves as a validation of the FDTD simulations shown here, as $\theta, \gamma_\parallel$
enter into the equations in a non-trivial manner from which the scaling relations are not apparent. 

In practice however checking these scaling relation may be difficult, 
as the total relaxation rate of the inversion, $\gamma_\parallel$,
can be written as a sum of contributions from spontaneous emission and non-radiative decay,
\begin{equation}
\gamma_\parallel = \gamma_{spon} + \gamma_{nr},
\end{equation}
in which the spontaneous decay rate can be written as \cite{shankar},
\begin{equation}
\gamma_{spon} = \frac{4 \alpha_{fs} \omega_a^3 n \theta^2}{3 c^2},
\end{equation}
where $\alpha_{fs}$ is the fine structure constant and $\gamma_{spon}$ is seen to be exactly dependent upon $\theta^2$. Thus, in the limit of an atomic gain media
without a non-radiative decay channel available from the upper level to the ground state,
the ratio of $\gamma_\parallel / \theta^2$ in the linewidth does not yield any new information
as these two parameters are not independent. However, this analysis does verify the intuitive
statement that the laser linewidth will be reduced if the non-radiative decay rate is substantially
larger than the spontaneous emission decay rate, decreasing the overall significance of spontaneous
emission to the system, as the relative ratio of $\theta^2 / \gamma_\parallel$ that
appears in Eq.~(\ref{CMTlw}) will be reduced.

As noted, the scaling of the linewidth with the ratio $\gamma_\perp \gamma_\parallel / \theta^2$ is not exact
as there is an additional dependence on $\gamma_\perp$ in the bad cavity factor once $\gamma_c \sim \gamma_\perp$.
In the right panel of Fig.~\ref{fig:scale}, the ratio
of $ \gamma_\perp \gamma_\parallel / \theta^2$ is held constant and equal to that in the left panel of the figure, 
but $\gamma_\perp$ is decreased so as to make the bad-cavity factor significantly different from unity. Thus instead
of remaining constant the linewidth decreases, in this case by roughly a factor of $2/3$.   
However it is possible to account for this failure of scaling by including a further approximate scaling 
by noting that when $\omega_a \approx \omega_0 \gg \gamma_\perp$,
we can express the bad-cavity factor as
\begin{equation}
B = \frac{1}{\left|\int \boldsymbol{\psi}_0^2(\mathbf{x}) \left(\varepsilon + \frac{\omega_0}{2} \frac{d\varepsilon}{d\omega}|_{\omega_0} \right) d\mathbf{x} \right|} \approx \left|\frac{1}{1+ \frac{\gamma_c}{2\gamma_\perp}}\right|. \label{eq:bcapprox}
\end{equation}
Using this, we can rescale the N-SALT linewidth prediction by $B_{new}^2/B_{old}^2$ calculated using the 
simple form on the right-hand side of Eq.~(\ref{eq:bcapprox}) (magenta dashed line).  With the additional rescaling
the N-SALT linewidth for the parameters of the left panel now 
agrees with the N-SALT prediction for the new gain media parameters in the right panel (green line) and
quantitatively agree with the FDTD simulations (cyan squares). This also verifies that the N-SALT form
of the bad-cavity factor correctly reduces to previously known approximations \cite{lax_conf_1966,kuppens_quantum-limited_1994,vanexter_theory_1995}, 
at least for the parameters chosen here.

\subsection{relaxation oscillation sidebands}

In Class B lasers, fluctuations in the amplitude of the electric field undergo relaxation
oscillations while decaying to the steady-state. These relaxation oscillations give rise to
side-peaks in the spectrum of the output intensity and in this section we will demonstrate that
the N-SALT is able to correctly reproduce the location and size of these side-peaks \cite{pick_linewidth_2015}.
It has been known for many decades that the relaxation oscillation frequency increases
as the laser is pumped further above threshold \cite{arecchi_ABC_84}, but
previous studies did not take into account the spatial variation in the gain saturation,
which was shown \cite{pick_linewidth_2015} to play an important role in quantitatively predicting the laser linewidth
in Sec.~\ref{sec:results}. Using the spatial lasing mode profiles and inversion calculated
using SALT, N-SALT demonstrates that the output intensity spectrum is dependent upon
the total local decay rate \cite{pick_linewidth_2015},
\begin{equation}
\gamma(\mathbf{x}) = \gamma_\parallel \left(1 + \frac{\gamma_\perp^2}{(\omega_0-\omega_a)^2 + \gamma_\perp^2}|\mathbf{E}_{\textrm{SALT}}(\mathbf{x})|^2 \right), \label{eq:gammaOfX}
\end{equation}
which contains contributions from both the non-radiative decay rate of the inversion, $\gamma_\parallel$,
as well as the local rate of stimulated emission given by the second term in Eq.~(\ref{eq:gammaOfX}).
N-SALT yields two main results for the effects of relaxation oscillations on the linewidth.
First, that relaxation oscillation side peaks will appear for cavities whose parameters
satisfy the inequality $\dwPick \ll \gamma_\parallel \ll \int A(\mathbf{x}) d\mathbf{x}$, in which 
\begin{equation}
A(\mathbf{x}) = 2 I \re\left[ \frac{i \omega_0 \boldsymbol{\psi}_0^2(\mathbf{x}) \frac{\partial \varepsilon(\omega_0)}{\partial I}}{2 \int \boldsymbol{\psi}_0^2(\mathbf{x}) \left(\varepsilon + \frac{\omega_0}{2} \frac{d\varepsilon}{d\omega}|_{\omega_0} \right) d\mathbf{x}}\right],
\end{equation}
where $I$ is the intensity of the electric field, as defined above.

Second, N-SALT
gives an explicit form for the output intensity spectrum in the presence of relaxation oscillations (with $\alpha = 0$):
\begin{gather}
S_{\textrm{N-SALT}}(\omega) = \frac{\dwPick}{\omega^2 + \left(\frac{\dwPick}{2}\right)^2} + 
\frac{\dwPick}{\omega^2 (1 - R(\omega))^2 + \tilde{R}(\omega)^2}, \label{eq:spec} \\
R(\omega) =  \int \frac{A(\mathbf{x})\gamma(\mathbf{x})}{\omega^2 + \left(\frac{\dwPick}{2} + \gamma(\mathbf{x})\right)^2} d\mathbf{x}, \\
\tilde{R} (\omega) = \int \frac{A(\mathbf{x})\gamma(\mathbf{x})\left(\frac{\dwPick}{2} + \gamma(\mathbf{x})\right)}{\omega^2 + \left(\frac{\dwPick}{2} + \gamma(\mathbf{x})\right)^2} d\mathbf{x}.
\end{gather}

\begin{figure}
\centering
\includegraphics[width=0.99\textwidth]{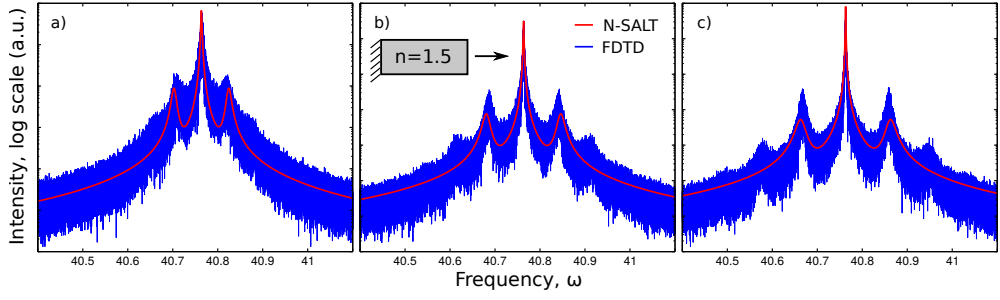}
\caption{Plots showing a comparison between the N-SALT prediction (red) and FDTD simulations (blue) of
the intensity spectrum for increasing values of the pump, $D_0$, for a single-sided, dielectric slab cavity with $n=1.5$,
$\omega_a = 40.7$, $\gamma_\perp = 1$, $\gamma_\parallel = 0.0025$, $\theta = 6 \times 10^{-10}$, and
$N_A = 10^{10}$. (a) $D_0 = 0.18$, (b) $D_0 = 0.28$, (c) $D_0 = 0.38$. As can be seen, increasing the
pump value increases the rate of stimulated emission, increasing $\gamma(x)$, Eq.~\ref{eq:gammaOfX},
resulting in increasing separation between the relaxation oscillation side peaks and the central lasing frequency.
In all three panels
of Fig.~\ref{fig:relax_osc}, the central frequency, $\omega_0$, chosen to evaluate Eq.~(\ref{eq:spec})
is the central frequency found by the FDTD simulations.
Intensity is plotted on a log scale in arbitrary units, rates are given in units of $c/L$, and
the inversion and total number of atoms are given in SALT units of $4 \pi \theta^2/\hbar \gamma_\perp$.
\label{fig:relax_osc}}
\end{figure}

The second term in Eq. \ref{eq:spec} describes the side peaks due to relaxation oscillations.
In Fig.~\ref{fig:relax_osc} we show the output intensity spectrum of a dielectric slab cavity pumped above
the first lasing threshold, in the parameter regime where side peaks are expected. Each of the
plots shows a comparison between the N-SALT prediction (red line) and the FDTD simulations (blue line)
for increasing values of the pump, (a) to (c). As can be seen in all three plots, excellent quantitative
agreement is seen between the simulated spectrum and the N-SALT prediction. To reiterate, N-SALT
has no free parameters, so the agreement seen here is a demonstration of a first principles test
of N-SALT. As can be see in the FDTD simulations, there are additional peaks in the
spectrum at a distance of twice the relaxation oscillation frequency from the central peak. In principle
N-SALT can be used to predict these additional side-peaks as well.
Finally, relaxation oscillations are proportional to the square root of the decay rate of the cavity,
$\omega_{\textrm{RO}} \sim \sqrt{(1/L)\int \gamma(\mathbf{x}) d\mathbf{x}}$, thus we expect for the side peaks seen in the
spectrum to move away from the central peak as the rate of stimulated emission increases
due to an increasing pump. As the pump is increased from Fig.~\ref{fig:relax_osc}(a) to
Fig.~\ref{fig:relax_osc}(c) we observe exactly this behavior in both the FDTD simulations
and N-SALT results, verifying this prediction.

\subsection{large alpha factor}

The $\alpha$ factor accounts for the phase fluctuations due to changes in the
susceptibility of the gain medium from intensity fluctuations, and is known to be quite
large in semiconductor gain material, where it is referred to as the Henry $\alpha$ factor. 
The N-SALT linewidth theory is quite general in
its derivation, and can be used to predict the linewidth of semiconductor lasers given
the appropriate form of the electric susceptibility. However, implementing an FDTD simulation
algorithm appropriate for semiconductor gain media is challenging even in the absence of
the effects of stimulated emission \cite{ho06,boehringer_part1_2008,boehringer_part2_2008,ravi_highly_2012,buschlinger_light-matter_2015}. 
Here, we test the N-SALT linewidth predictions
using two-level atomic gain media; the appropriate $\alpha$ factor in this case was first derived by Lax as \cite{lax_conf_1966},
\begin{equation}
\alpha_0 = \frac{\omega_0 - \omega_a}{\gamma_\perp}.
\end{equation}
For the simulations here, we choose the atomic transition frequency almost exactly in between
the two proximal cavity resonances, and decrease $\gamma_\perp$, thus increasing $\alpha_0$.
However N-SALT predicts a generalized $\alpha$ factor, $\tilde{\alpha}$,  which is sensitive to the
spatial hole-burning of the gain medium and the non-Hermitian nature of the lasing mode, and not
just to the distance of the lasing mode from the center of the gain curve.  From ref.~\cite{pick_linewidth_2015} in the single mode 
case it take the form:
\begin{equation}
\tilde \alpha = \frac{\im[C_{11}]}{\re[C_{11}]},
\end{equation}
where the complex coefficients $C_{\mu \nu}$ determine the relaxation rate of modal fluctuations away from
the steady-state lasing values.  These coefficients are calculated from the SALT solutions according to:
\begin{equation}
C_{\mu \nu} = \left[ \frac{i \omega_\mu \int \boldsymbol{\psi}_\mu^2(\mathbf{x}) \frac{\partial \varepsilon(\omega_\mu)}{\partial I_\nu} d\mathbf{x}}{2 \int \boldsymbol{\psi}_\mu^2(\mathbf{x}) 
\left(\varepsilon + \frac{\omega_\mu}{2} \frac{d\varepsilon}{d\omega}|_{\omega_\mu} \right) d\mathbf{x}}\right],
\end{equation}
where $\boldsymbol{\psi}_\mu(\mathbf{x})$ is the spatial profile of the $\mu$th lasing mode,
still power normalized, $\int \boldsymbol{\psi}_\mu^2(\mathbf{x}) d\mathbf{x} = 1$.
Furthermore, it was found in Sec.~7A-B in Pick \textit{et al.}\ \cite{pick_linewidth_2015}, that the
spatial profile of the first threshold lasing mode changes discontinuously as the passive
cavity dielectric constant is increased, jumping when the first lasing mode switches from one passive cavity
resonance to the next as different resonances enter and leave the bandwidth of the gain medium.
Near these discontinuities, a large deviation between $\alpha_0$ and $\tilde \alpha$ can
be observed, and we will exploit this phenomenon in the simulations below while maintaining
an index of refraction similar to that of GaAs, using $n=3.5$ for the dielectric slab cavity
studied here.

\begin{figure}
\centering
\includegraphics[width=0.60\textwidth]{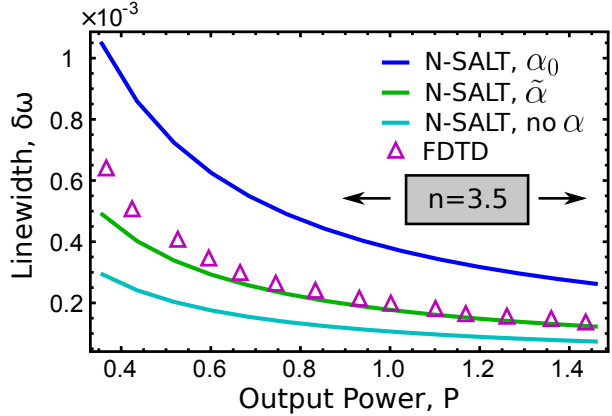}
\caption{Plot of the linewidth versus the output power for a two-sided dielectric
slab cavity, $n=3.5$, showing the comparison between the N-SALT linewidth prediction (green line),
the N-SALT linewidth without an $\alpha$ factor (cyan line), the N-SALT linewidth using Lax's 
$\alpha$ factor (blue line), and the FDTD simulation results (magenta triangles). Excellent
quantitative agreement is seen between the FDTD simulations and the correct N-SALT linewidth
prediction, confirming the form of the $\alpha$ factor derived by Pick \textit{et al.}\ \cite{pick_linewidth_2015}.
For the two-level gain medium used here, $\omega_a = 18.3$, $\gamma_\perp = 0.05$, $\gamma_\parallel = 0.01$,
$\theta = 4 \times 10^{-9}$, and $N_A = 10^{10}$, and results in the total system having
$\alpha_0^2 = 2.56$, while $\tilde \alpha^2 \approx 0.66$. Frequencies and rates are given in units of $c/L$, while the
atomic values are given in SALT units of $4 \pi \theta^2/\hbar \gamma_\perp$.
\label{fig:alpha}}
\end{figure}

In Fig.~\ref{fig:alpha} we show the results of a comparison between the N-SALT linewidth
predictions using three different $\alpha$ factors, $\alpha^2=0$ (cyan line), $\tilde \alpha^2 \approx 0.66$ (green line), and $\alpha_0^2 = 2.56$ (blue line),  
with direct FDTD simulation (magenta triangles). We find excellent agreement between
the correct N-SALT linewidth calculated using $\tilde \alpha$ and the FDTD simulations,
demonstrating that this is the correct form of the $\alpha$ factor. These simulations
also verify that the Langevin noise model used in the FDTD simulations implicitly contains
the physical effects that yield the $\alpha$ factor. While the $\alpha$ factor for 
many semiconductor lasing materials is determined experimentally \cite{osinski_linewidth_1987}, rather than analytically,
these results indicate that the physical origins of the phenomenon are effected by the
geometry of the cavity and the spatial profile of the lasing mode. Furthermore, this suggests
that using fabrication techniques to control the index of refraction of semiconductor
based laser cavities should allow for the engineering of different linewidth enhancement
factors.

\subsection{two mode lasing}

A final feature of N-SALT linewidth theory is that it predicts the linewidths in 
the multimode steady state, and finds that the linewidths are not independent of one another
but couple through gain saturation.  Specifically it predicts that the linewidth of any active lasing
mode is affected by the onset of additional lasing modes at higher pump powers.
This coupling phenomenon occurs through a change in the
$\alpha$ factor of each active mode at each subsequent threshold. Above the second lasing threshold this correction is 
given by
\begin{equation}
\dwPick^{(two-mode)} = \dwPick^{(1)} \left[1 + \frac{C_{11}^I C_{22}^R - C_{21}^I C_{21}^R}{C_{11}^R C_{22}^R - C_{12}^R C_{21}^R}\right] + \dwPick^{(2)}\left[\frac{C_{11}^R C_{12}^I - C_{11}^I C_{12}^R}{C_{11}^R C_{22}^R - C_{12}^R C_{21}^R}\right],
\label{eq:two_mode}
\end{equation}
in which $\dwPick^{(i)}$ is the single-mode N-SALT linewidth prediction from Eq.~\ref{eq:noiseFDTDCMT},
and the superscripts $R$ and $I$ denote the real and imaginary components of the amplitude relaxation
rates $C_{ij}$ respectively. Near threshold this analytic expression diverges and is not valid, but N-SALT
dynamical equations can be integrated numerically to calculate the increase in the linewidth due to the second
lasing mode \cite{pick_linewidth_2015}.

\begin{figure}
\centering
\includegraphics[width=0.60\textwidth]{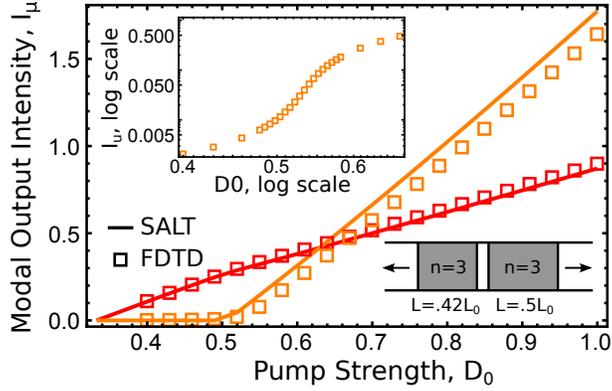}
\caption{Plot showing the modal output intensity as a function of the gain medium
pump strength $D_0$, for a two-sided system consisting of two coupled dielectric cavities, $n=3$, with different
lengths, $L_1 = .42L_0$, and $L_2 = .5L_0$, joined together by a region of air, $n=1$, with length
$L_{air}=.08L_0$, where $L_0$ is the total size of the system, and as shown in the schematic. 
This cavity has up to two active lasing modes (red and orange) for the pump values simulated here, and
quantitative agreement is seen between the SALT simulations (solid lines) and noisy FDTD simulations (squares).
A slight offset in the interacting threshold for the second lasing mode is seen between the two simulations,
with $D^{(2)}_{SALT} = 0.5077$, while $D^{(2)}_{FDTD} = 0.5282$. The inset plot shows the FDTD simulated intensity
of the second lasing mode through its lasing threshold, first showing amplified spontaneous emission, then
super-linear behavior at threshold, and finally linear behavior above threshold, as expected.
The gain medium was chosen
to have $\omega_a = 15$, $\gamma_\perp = 0.4$, $\gamma_\parallel = 0.01$, $\theta = 10^{-9}$,
and $N_A = 10^{10}$. Frequencies and rates are given in units of $c/L_0$, while the
field quantities and inversion values are given in SALT units of $4 \theta^2/\hbar^2 \gamma_\perp \gamma_\parallel$ and $4 \pi \theta^2/\hbar \gamma_\perp$, respectively.
\label{fig:2mode_inten}}
\end{figure}

\begin{figure}
\centering
\includegraphics[width=0.60\textwidth]{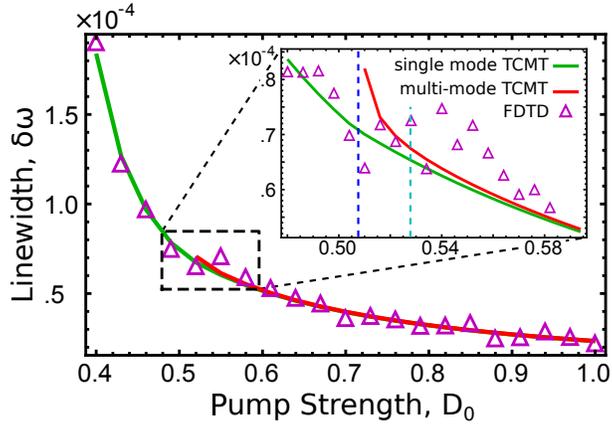}
\caption{Comparison of the single mode N-SALT linewidth prediction (green line), two-mode N-SALT linewidth
prediction (red line), and FDTD simulations (magenta triangles) for the first lasing mode in coupled cavity system from Fig.~\ref{fig:2mode_inten}.
Inset shows a zoom in of the same quantities close to the interacting threshold of the second lasing mode.
The two slightly different second mode thresholds are marked in the inset, $D^{(2)}_{SALT}$ (dashed blue line), and $D^{(2)}_{FDTD}$ (dashed cyan line).
While the data is too noisy, and the difference between the single mode and two-mode predictions too small,
for the resolution of their differences, we do observe increased linewidth and variance in our simulations
close to the threshold of the second lasing mode, as expected.
\label{fig:2mode_mode1}}
\end{figure}

To study this effect, we used two coupled
dielectric cavities as shown in the schematic of Fig.~\ref{fig:2mode_inten}, with the total system
open on both ends. By using coupled cavities we create doublets of resonances, and by then placing
the gain frequency center near one doublet we restrict ourselves to two mode lasing, which is convenient for
the FDTD simulations in particular.
The semiclassical prediction for the modal intensities as a function of the pump
strength for this cavity calculated using SALT (solid lines) is shown in Fig.~\ref{fig:2mode_inten},
and compared against the FDTD simulations (squares), demonstrating quantitative agreement. The inset
plot shows the super-linear behavior through the lasing threshold observed in the FDTD simulations,
as the amplified spontaneous emission yields a coherent lasing signal. We observe excellent
quantitative agreement between the N-SALT prediction and the FDTD simulations for the linewidth
of the first lasing mode, as shown in Fig.~\ref{fig:2mode_mode1}. On this scale, the single-mode
N-SALT prediction (green) is very similar to the multi-mode prediction, Eq.~\ref{eq:two_mode} (red).
However, the inset of Fig.~\ref{fig:2mode_mode1} shows the same set of comparisons through the
turn on of the second lasing mode. Unfortunately, while there is clearly enhanced noise near the 
second modal threshold, there is not enough resolution in the FDTD data to compare accurately
the single-mode and two-mode N-SALT predictions.
There are two difficulties with the numerical comparison. 
First, the design of the system, and the heirarchy of parameter scales that must be achieved
above the floor of the spectral resolution of the simulation results in a noisy signal. Second,
due to discretization errors, the SALT and FDTD simulations give slightly different predictions
for the location of the second lasing threshold. Thus, when plotted against the pump strength,
we expect the linewidth increase in the FDTD simulations to occur at a slightly shifted location
relative to the N-SALT results, but the noise makes it difficult to extract this shift.

\begin{figure}
\centering
\includegraphics[width=0.60\textwidth]{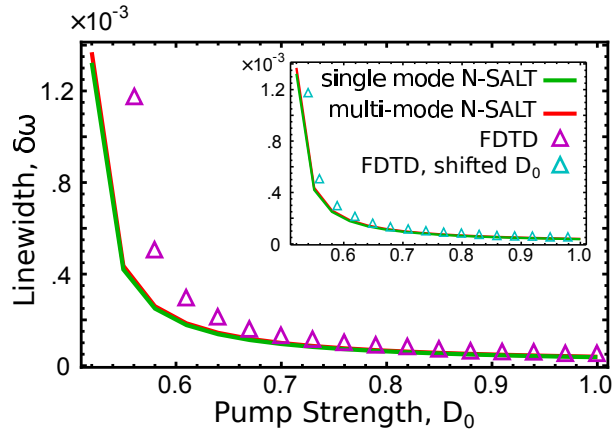}
\caption{Comparison of the single mode N-SALT linewidth prediction (green line), two-mode N-SALT linewidth
prediction (red line), and FDTD simulations (magenta triangles) for the second lasing mode in coupled cavity system from Fig.~\ref{fig:2mode_inten}.
Inset shows the same data except with the FDTD simulations plotted at shifted pump values (cyan triangles) to account
for the slightly different second lasing mode thresholds seen in Fig.~\ref{fig:2mode_inten}. Quantitative
agreement between the FDTD and N-SALT linewidth predictions is seen in both versions of the plot, but the
inset demonstrates that most of the discrepancy seen in the outer plot is due to differences in the output
power of the cavity due to the SALT simulations being further above threshold than the FDTD simulations
for the same value of the pump $D_0$.
\label{fig:2mode_mode2}}
\end{figure}

Although the FDTD simulations are not sensitive enough to observe the small corrections in the first mode
linewidth due to onset of the second mode, these simulations are able to validate the N-SALT linewidth prediction
for the second lasing mode. Figure \ref{fig:2mode_mode2} shows a comparison between the N-SALT
prediction and FDTD simulations for the linewidth of the second lasing mode as a function of the
input pump strength. 
As noted, the offset observed between the two linewidths is due to the slightly different
locations of the second mode threshold and if this difference is subtracted, as is seen in
the inset of Fig.~\ref{fig:2mode_mode2}, we see excellent quantitative agreement between the
two sets of simulations.

\section{Summary \label{sec:summary}}

In this work we have performed a first principles test of the N-SALT linewidth
results derived by Pick \textit{et al.}\ \cite{pick_linewidth_2015}. To do this, we used the FDTD
algorithm to simulate the Maxwell-Bloch equations coupled to a set of Langevin noise
equations, thus including the effects of spontaneous emission. We found excellent quantitative
agreement between the N-SALT linewidth predictions and the FDTD simulations, while finding
substantial deviations from the `fully corrected' Schawlow-Townes theory, demonstrating that
the intertwining of the cavity decay rate, Petermann factor, incomplete inversion factor, bad-cavity correction
and Henry $\alpha$ factor in the N-SALT linewidth formula is necessary and correct. This comparison was first
done in a parameter range in which the relaxation oscillations were weak (near the Class A boundary).
Through comparison with the Chong and Stone linewidth theory \cite{chong12}, we demonstrated
that for the small, $20 \lambda_a \sim L$, cavities studied here, much but not all of the improved agreement 
found by N-SALT is due to the proper treatment of the incomplete inversion factor. Next, we
successfully demonstrated that N-SALT gives the correct output intensity spectrum including relaxation oscillations
for Class B lasers, and correctly reproduces the side-peaks due to relaxation oscillations.
This set of simulations also verified that the side-peaks shift away from the center of the
spectrum as the pump on the gain medium is increased. We then studied the different predictions
for the linewidth enhancement due to the coupling between intensity and phase fluctuations, the $\alpha$ factor,
and demonstrated that the N-SALT form of the $\alpha$ factor yields quantitative agreement
with the FDTD simulations, while previous forms of the $\alpha$ factor are shown to disagree.
This set of simulations is particularly remarkable, because in the absence of the N-SALT prediction
for $\tilde \alpha$, one might conclude that the FDTD simulations do not correctly capture
the effects of the $\alpha$ factor. Instead, it is clear that
the FDTD algorithm used does contain all of the relevant physics, and that there can be
a significant difference between the various forms of the $\alpha$ factor. Finally,
we demonstrated that the N-SALT theory correctly predicts the linewidth for multiple
active lasing modes.

\section*{Acknowledgments}
We thank Alejandro Rodriguez, Bradley Hayes, Arthur Goestchy, and Jonathan Andreasen for helpful discussions. 
This work was supported by NSF grant
No.~DMR-1307632. 
A.P.~and S.G.J.~acknowledge support from
Army Research
Office through the Institute for Soldier Nanotechnologies
under Contract No.~W911NF-13-D-0001.
C.Y.D.~acknowledges support by the Singapore National
Research Foundation under grant No.~NRFF2012-02. 
This work was supported in part by the facilities and staff of the Yale 
University Faculty of Arts and Sciences High Performance Computing Center.  

\end{document}